\documentclass{aa}
\newcommand{\hii}{H{\sc ii}}
\newcommand{\mia}{W3\,IRS5}
\newcommand{\rxa}{1.4\,mm}
\newcommand{\rxb}{3.4\,mm}
\newcommand{\so}{SO$_2$}
\newcommand{\solar}{$_{\odot}$}
\newcommand{\soa}{\mbox{\so{ (22$_{2,20}$--22$_{1,21}$)}}}
\newcommand{\sob}{\mbox{\so{ (8$_{3,5}$--9$_{2,8}$)}}}
\newcommand{\sioa}{\mbox{SiO~(5--4)}}
\newcommand{\siob}{\mbox{SiO~(2--1)}}

\usepackage{graphicx}
\usepackage{natbib}
\usepackage{txfonts}
\begin{document}
   \title{Millimeter interferometry of \mia{}\thanks{FITS images are available in electronic form at the CDS via anonymous ftp to {\tt cdsarc.u-strasbg.fr (130.79.128.5)} or via {\tt http://cdsweb.u-strasbg.fr/cgi-bin/qcat?J/A+A/}}: A Trapezium in the making}

   \author{J. A. Rod\'on\inst{1}
          \and
          H. Beuther\inst{1}
	  \and
	  T. Megeath\inst{2}
	  \and
	  F. F. S. van der Tak\inst{3}
          }

   \institute{Max-Planck-Institut f\"ur Astronomie,
              K\"onigstuhl 17, 69117 Heidelberg, Germany.\\
              \email{rodon@mpia.de}
         \and
		Department of Physics, University of Toledo, 2801 W. Bancroft St, Toledo OH 43606, USA.
	\and
		SRON Netherlands Institute for Space Research, 
		Landleven 12, 9747 AD Groningen, The Netherlands.
             }


 
  \abstract
   {Although most young massive stars appear to be part of multiple systems, it is poorly understood how this multiplicity influences the formation of massive stars. The high-mass star-forming region \mia{} is a prime example of a young massive cluster where the cluster center is resolved into multiple subsources at cm and infrared wavelengths, a potential proto-Trapezium system.}
   {We investigate the protostellar content in the \rxa{} continuum down to subarcsecond scales and study the compact outflow components, also tracing the outflows back to their driving sources via the shocktracing SiO and \so{} emission.}
   {The region \mia{} was mapped with the PdBI at \rxa{} and \rxb{} in the AB configurations, tuning the receivers to observe the molecular transitions \soa{}, \sob{}, \siob{}, and \sioa{}.}
   {In the continuum we detect five sources, one of them for the first time, while counterparts were detected in the NIR, MIR or at radio wavelengths for the remaining four sources. Three of the detected sources are within the inner 2100\,AU, where the protostellar number density exceeds $10^6$ protostars\,pc$^{-3}$ assuming spherical symmetry. Lower limits for the circumstellar masses of the detected sources were calculated, ranging from $\sim$0.3 to $\sim$40\,M\solar{} although they were strongly affected by the spatial filtering of the interferometer, losing up to $\sim$90$\%$ of the single-dish flux. However, the projected separations of the sources ranging between $\sim750$ and $\sim4700$ AU indicate a multiple, Trapezium-like system. We disentangled the compact outflow component of \mia{}, detecting five molecular outflows in SiO, two of them nearly in the line of sight direction, which  allowed us to see the collapsing protostars in the NIR through the cavities carved by the outflows. The \so{} velocity structure indicates a rotating, bound system, and we find tentative signatures of converging flows as predicted by the gravoturbulent star formation and converging flow theories.}
   {The obtained data strongly indicate that the clustered environment has a major influence on the formation of high-mass stars; however, our data do not clearly allow us to distinguish whether the ongoing star-forming process follows a monolithic collapse or a competitive accretion mechanism.} 


   \maketitle
%

\section{Introduction}

The formation of high-mass stars appears to be intimately linked with the formation of multiple stellar systems. Optically visible OB-type stars show a much higher degree of multiplicity than their lower-mass counterparts \citep{preibisch1999}. Now observations are showing that such high-mass multiple systems have separations ranging between 10,000\,AU and 1\,AU, and appear to be bound groups within larger clusters \citep{mermilliod2001}. In contrast to hierarchical systems where the successive separation between its members increases by large factors, many of these systems are dynamically unstable, nonhierarchical clusters of three or more stars, called trapezia after the Trapezium cluster in Orion. The role such systems play in massive star formation is not yet understood, although several suggestions have been made.

First, the mass density of these multiple systems may be high enough to gravitationally trap hypercompact or even small ultracompact \hii{} regions. \citet{keto2002a,keto2002b,keto2003} has shown that continued accretion through the HC\hii{} region may be possible, even after the supporting stars have reached the main sequence.
The radius where these \hii{} regions become hydrodynamically supported -thus stopping the accretion flow- scales inversely with the attracting mass, thus massive trapezia systems allow more mass to be accreted to the central accreting objects \citep{keto2002a,keto2002b,keto2003}.

Second, binary systems are expected to form in the center of these dense proto-Trapezia. These binaries will continue accreting gas, funneled to the center by the combined potential of the protocluster. Because the angular momentum of the infalling gas has no correlation with that of the binary's, the masses of the individual stars increase while the separation of the binary decreases, giving as a natural outcome a high-mass close binary system that could eventually undergo mergers, creating even more massive systems \citep{bonnell2005}.

The previous examples show how the formation of massive stars in multiple systems with separations on the order of $\sim$1000\,AU may be affected by their companions, either by modifying the morphology of accreting envelopes allowing more mass to be accreted
, or by forming more massive stars by merging high-mass close binaries.

Some multiple high-mass protostellar systems are known. For example, W\,33A was imaged by \citet{vandertak2005b} at 43\,GHz with the Very Large Array finding three continuum sources at separations of $3000-5000$\,AU; \citet{hunter2006} found similar systems in NGC\,6334\,I and NGC\,6334\,I(N) with a few millimetric sources inside a region of $\sim$10,000\,AU. More recently, \citet{beuther2007d} resolved the central 7,800\,AU of the hot molecular core G29.96--0.02 into four submillimeter peaks, a potential proto-Trapezium system.

One prime example of a trapezia is the system \mia{} in the W3-Main region of the \hii{}/molecular cloud complex W3, a very active star-forming region part of the larger star-forming complex W3/W4 located in the Perseus arm of the Galaxy, that also encompasses the W4 star-forming region and the OB associations IC 1805 and IC 1795. In W3 the star formation takes place in the embedded regions W3-Main, W3-North, and W3-OH.

\citet{oey2005} found evidence supporting the scenario in which the star formation in W3-Main was triggered by the formation of IC 1795, which in turn was triggered by the Perseus chimney/superbubble. This scenario, however, has been brought into question by the recent Chandra X-Ray observations of \citet{feigelson2008}, because W3-Main does not show the elongated and patchy structure of a triggered star cluster.
W3-Main itself may harbor its own triggered star-formation scenario because the various morphological classes of \hii{} regions that can be found in it suggest that different stages of massive star formation (MSF) may be sequentially triggered by the pressure of the expanding \hii{} regions \citep{tieftrunk1997,tieftrunk1998}.

The system \mia{} is a known double infrared (IR) source \citep{howell1981,neugebauer1982} with a total luminosity of 2\,$\times$\,10$^5\,L_{\odot{}}$ \citep{campbell1995}, located in the W3-Main region at a distance of 2\,kpc \citep{megeath2008}. Recently, \citet{megeath2005} presented HST observations with an angular resolution of 350\,AU identifying seven near-IR sources including the two previously known. Three of these sources have both mid-IR counterparts and are 1.2 and 0.7\,cm continuum sources \citep{wilson2003,vandertak2005a}. The sizes of the \hii{} regions are $<$240\,AU, indicating that they are probably gravitationally bound and may be accreting \citep[e.g.,][]{keto2003}. Based on its size, the probable masses of the stars, and its nonhierarchical distribution of sources, \citet{megeath2005} proposed that this system is a proto-Trapezium system, which would emerge as a bound Trapezium similar to that in the Orion nebula.
This region is also the source of at least two outflows \citep{imai2000,wilson2003}, and is in the center of an embedded cluster of 80--240 low-mass stars \citep{megeath1996} which in turn is surrounded by a core of several hundred low-mass stars \citep{feigelson2008}.

\section{Observations}
We have mapped \mia{} with the IRAM \textit{Plateau de Bure Interferometer} (PdBI)\footnote{IRAM is supported by INSU/CNRS (France), MPG (Germany) and IGN (Spain).} in the \textit{A} (Jan/Feb 2006) and \textit{B} (Mar 2006) configurations at 1.4 and 3.4\,mm imaging the continuum, achieving an angular resolution with a synthesized beam of 0.39\arcsec{}\,$\times$\,0.34\arcsec{} at 1.4\,mm and 1.00\arcsec{}\,$\times$\,0.89\arcsec{} at 3.4\,mm. At the given distance of 2\,kpc that translates to a spacial resolution of $\sim$700 and $\sim$1900\,AU, respectively. The $\sim 0.36\arcsec{}$ resolution obtained was feasible because of the new very extended baselines, extending to over 700\,m.

The receivers were tuned to 87 and 217\,GHz both in the lower sideband. With this spectral setup we observed the SO$_2$, SiO and C$^{18}$O transitions described in Table \ref{table-lines} with a maximum spectral resolution of 0.5\,km\,s$^{-1}$. We adopted a systemic velocity V$_{\mathrm{LSR}}=-39$\,km\,s$^{-1}$ \citep{ridge2001,tieftrunk1998}. Although observed, C$^{18}$O was not detected due to the short-spacings problem inherent to the interferometer filtering out the received flux.

\begin{table}[h]
\renewcommand{\arraystretch}{1.2}  
\begin{minipage}[]{\columnwidth}
 \caption{Observed molecular transitions and rms of the respective maps.}
\label{table-lines}
\centering
\renewcommand{\footnoterule}{}
\begin{tabular}{lcccc}
 \hline\hline
& $\nu$ & Spect. Resol. & E$_{up}$ & rms\footnote{For a spectral resolution of 0.5\,km\,s$^{-1}$ except for \sioa{}, whose map has a spectral resolution of 1\,km\,s$^{-1}$.}\\
Transition & (GHz) & (MHz) & (K) & (mJy beam$^{-1}$)\\
\hline
\sob{} & 86.64 & 0.14 & 55 & 8.0 \\
\siob{} & 86.85 & 0.14 & 6 & 7.0\\
Continuum & 87.02 & \ldots & \ldots & 1.8 \\
\soa{} & 216.64 & 0.36 & 248 & 18.0 \\
Continuum & 216.87 & \ldots & \ldots & 1.2 \\
\sioa{} & 217.10 & 0.36 & 31 & 14.0 \\
C$^{18}$O(2--1)\footnote{Not detected, filtered out by the interferometer.} & 219.56 & \ldots & 15 & \ldots \\
\hline\hline
\end{tabular}
\end{minipage}
\end{table}

The phase and amplitude calibrators were 0355+508 and 0059+581 and the flux calibrator was 3C345, adopting the values from the SMA flux monitoring of these quasars. The data were calibrated with the program CLIC and then imaged with the program MAPPING, both part of the GILDAS package. The spectra were processed with the program CLASS77, also from the GILDAS package.

After inversion and cleaning, the continuum data have rms noise levels $\sigma=1.8$\,mJy\,beam$^{-1}$ and $\sigma=1.2$\,mJy\,beam$^{-1}$ at \rxb{} and \rxa{} respectively. For the line data, with a spectral resolution of 0.5\,km\,s$^{-1}$ the noise levels are $\sigma=7$\,mJy\,beam$^{-1}$ for \siob{}, $\sigma=18$\,mJy\,beam$^{-1}$ for \soa{} and $\sigma=8$\,mJy\,beam$^{-1}$ for \sob{}, while \sioa{} was imaged with a spectral resolution of 1\,km\,s$^{-1}$ resulting in a rms $\sigma=14$\,mJy\,beam$^{-1}$ (see Table \ref{table-lines}).

\begin{table*}[t!]
\renewcommand{\arraystretch}{1.2}  
\begin{minipage}[]{2\columnwidth}
	\centering
 	\caption{Properties of the continuum millimetric sources in \mia{}.}
\label{table-sources}
\begin{tabular}{lcc|cc|ccc|cc}
\hline\hline
 & R.A. & Dec. & I$_{\nu}$\footnote{The fluxes have been corrected for the free-free contribution, except for source MM4} & S$_{\nu}$\,$^a$ & \multicolumn{3}{c}{Mass (M\solar{})} & N$\left(\mathrm{H}_2\right)$\footnote{Calculated assuming $\mathrm{T}=100\mathrm{K}$.} & A$_{\textrm{\scriptsize v}}$ \\
MM & (J2000) & (J2000) & (mJy beam$^{-1}$) & (mJy) & 50\,K & 100\,K & 200\,K & ($10^{24}$ cm$^{-2}$) & (10$^3$ mag) \\ 
\hline
\multicolumn{10}{c}{\rxa{} with 0.39"\,$\times$\,0.34" beam} \\
\hline
1 \ldots & 02 25 40.77 & 62 05 52.49 & 34.8 & 70 & 6.7 & 3.2 & 1.6 & 8.0 & 8.5 \\
2 \ldots & 02 25 40.68 & 62 05 51.53 & 20.6 & 24 & 2.3 & 1.1 & 0.5 & 4.7 & 5 \\
3 \ldots & 02 25 40.66 & 62 05 51.95 & 10.4 & 13 & 1.2 & 0.6 & 0.3 & 2.4 & 2.5 \\
4 \ldots & 02 25 40.75 & 62 05 50.45 & 5.7\footnote{Free-free contribution not subtracted.} & 5.7\footnote{Unresolved, I$_{\nu}$ given instead of S$_{\nu}$.} & 0.6\footnote{Calculated with the peak intensity instead of the flux density, because the angular size of the source is smaller than the beam.} & 0.3$^e$ & 0.1$^e$ & 1.3 & 1.4 \\
\hline
\multicolumn{10}{c}{\rxb{} with 1.00"\,$\times$\,0.89" beam} \\
\hline
1 \ldots & 02 25 40.77 & 62 05 52.49 & 23.1 & 24 & 83.4 & 40.1 & 20.2 & 30 & 32 \\
5\footnote{Source MM5 is likely the merged pair MM2 and MM3.} \ldots & 02 25 40.68 & 62 05 51.85 & 11.2 & 11.2$^d$ & 39.0$^e$ & 19.1$^e$ & 9.4$^e$ & 14 & 15 \\
6 \ldots & 02 25 40.73 & 62 05 49.86 & 5.1 & 5.1$^d$ & 17.7$^e$ & 8.7$^e$ & 6.3$^e$ & 6.6 & 7 \\
\hline\hline
\end{tabular}
\renewcommand{\footnoterule}{} 
\end{minipage}
\end{table*}

\section{Results}
\subsection{Large-scale Overview}

The W3 complex is a very rich one when it comes to \hii{} regions. Several \hii{} regions have been detected at 6, 2 and 1.3\,cm by \citet{tieftrunk1997}, encompassing hypercompact, ultracompact, compact and diffuse \hii{} regions. All of them probably originated from the same volume of neutral gas.

With the ages they estimated, \citet{tieftrunk1997} suggest a qualitative evolutionary sequence of the \hii{} regions, going from hypercompact to ultracompact, compact and then diffuse. Also the number, age and spatial distribution of the \hii{} regions suggest a scenario of triggered star formation by the expansion of the \hii{} regions that formed first.

\begin{figure}[h]
	\centering
	\includegraphics[angle=-90,width=\columnwidth]{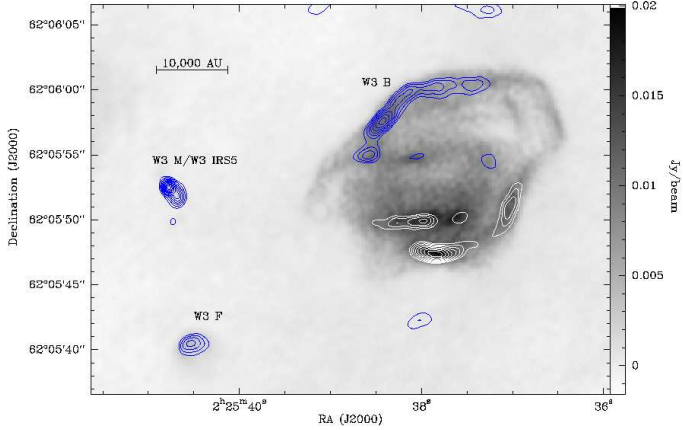}
	\caption{Continuum image (greyscale) of the W3 region at 4.9 GHz adapted from \citet{tieftrunk1997}. Contours trace the observed PdBI continuum emission at 3.4 mm, starting at the $3\sigma$ level in $1\sigma$ steps. Shown are the \hii{} regions W3 B (compact), W3 F (ultracompact) and W3 M/\mia{} (hypercompact). W3 B lies at the border of our primary beam, thus the contours above and below W3 B are artifacts of the deconvolution process, and do not have any physical meaning.}
	\label{large_scale}
\end{figure}

In our \rxb{} map we detected three of those \hii{} regions, the compact W3 B, the ultracompact W3 F and the hypercompact \mia{}, also known as W3 M, the subject of our study (see Fig. \ref{large_scale}). The shell-like structure and overall morphology of W3 B is recovered by our observations, as well as the cometary shape of W3 F.

\subsection{Millimetric Continuum}

Figure \ref{cont} shows the 1.4 and 3.4\,mm continuum maps, with the six sources detected labeled MM1--6 in descending order of intensity at 1.4\,mm for MM1--4 and at 3.4\,mm for MM5 and MM6. MM1 is the only source resolved at both wavelengths, while MM4 is detected only at 1.4\,mm and MM6 only at 3.4\,mm. At the long wavelength we do not resolve MM2 nor MM3. However, based on its position, MM5 is likely the joint contribution from MM2 and MM3 and is not considered an individual source.

The properties of the sources are summarized in Table \ref{table-sources}. Columns 2 and 3 give their positions; the measured peak flux intensity and integrated flux density, both corrected for the free-free contribution, are given in columns 4 and 5 respectively.

In the standard model of optically-thick (spectral index $\sim2$) and optically-thin (spectral index $\sim-0.1$) free-free emission, the spectral indexes derived by \citet{vandertak2005a} indicate that the 43\,GHz emission is close to the ``turnover point'' between optically-thick and optically-thin free-free emission, and is approximately an upper limit to the free-free contribution at millimetric wavelengths. In HC\hii{} region models with density gradients \citep[e.g.,][]{keto2003} or clumpiness \citep{sewilo2004} the free-free emission may rise somewhat more, but unlikely by a significant amount. Therefore, here we assume that the 43\,GHz emission is a good measure of the free-free contribution at millimetric wavelengths and we use it to correct our measured fluxes for such contribution, as shown in Table \ref{ffree}. This could be done for all our sources except MM4, for which \citet{vandertak2005a} did not detect a counterpart.

\begin{table}[h]
	\renewcommand{\arraystretch}{1.2}
\begin{minipage}[]{\columnwidth}
	\caption{Contribution of the free-free emission to the measured millimetric fluxes.}
	\label{ffree}
	\centering
\begin{tabular}{ccc|cc}
	\hline\hline
 & \multicolumn{2}{c}{Peak Intensity (f-f)} & \multicolumn{2}{c}{Flux Density (f-f)}\\
MM & (mJy beam$^{-1}$) & $\frac{\mathrm{free-free~I}_{\nu}\footnote{Estimated from \citet{vandertak2005a}, see discussion in the text.}}{\mathrm{measured~I}_{\nu}}$ & (mJy) & $\frac{\mathrm{free-free~S}_{\nu}\,^a}{\mathrm{measured~S}_{\nu}}$\\
\hline
\multicolumn{5}{c}{\rxa{} with 0.39"\,$\times$\,0.34" beam} \\
\hline
1 \ldots & 2 & 0.05 & 2 & 0.03\\
2 \ldots & 2 & 0.09 & 4 & 0.14\\
3 \ldots & 2 & 0.16 & 3 & 0.19\\
\hline
\multicolumn{5}{c}{\rxb{} with 1.00"\,$\times$\,0.89" beam} \\
\hline
1 \ldots & 2 & 0.08 & 2 & 0.08\\
5 \ldots & 4 & 0.26 & 4\footnote{Unresolved in our data, free-free emission correction for I$_{\nu}$ used.} & 0.26$^b$\\
6 \ldots & 0.7 & 0.12 & 0.7$^b$ & 0.12$^b$\\
\hline
\hline
\end{tabular}
\renewcommand{\footnoterule}{} 
\end{minipage}
\end{table}

Because the brightness temperature of the corresponding Planck function for the strongest source --MM1-- is about 11\,K, just $\sim$10$\%$ of the typical hot core temperatures of $\sim$100\,K, we can assume that the emission comes from optically thin dust and thus calculate the masses and column densities with the approach outlined by \citet{hildebrand1983} and adapted by \citet{beuther2002a,beuther2002erratum}. We adopted a distance of 2\,kpc \citep{megeath2008} and used a grain emissivity index $\beta = 2$, corresponding to a dust opacity per unit mass $\kappa_{3.4mm} \sim 0.05$ and $\kappa_{1.4mm} \sim 0.3$\,cm$^2$\,g$^{-1}$ for a median grain size $a = 0.1\,\mu$m and grain mass density $\rho = 3$\,g\,cm$^{-3}$. In columns 6--8 and 9 of Table \ref{table-sources} are the calculated masses and H$_2$ column densities respectively. We did not derive a temperature for this region, therefore the masses were calculated assuming the different temperatures T\,$\sim50$\,K, T\,$\sim100$\,K and T\,$\sim200$\,K, while for the N$\left(\mathrm{H}_2\right)$ calculation we assume a hot core temperature of $\sim100$\,K.

The obtained masses are strongly affected by the spatial filtering inherent to interferometers. Comparing with single-dish data \citep{vandertak2000b} we recover only $\sim 9\%$ of the flux at \rxa{} (see Sec. \ref{cont_disc} for further discussion).
The visual extinctions in column 10 were calculated assuming \mbox{A$_{\mathrm{v}} = \mathrm{N}\left(\mathrm{H}_2\right) / 0.94 \times 10^{21}$} \citep{frerking1982}.

\subsection{Silicon Monoxide emission}
\label{sio2}

Silicon Monoxide (SiO) is believed to trace strong shocks in dense molecular gas, particularly in molecular outflows \citep{schilke1997a}. Here two SiO transitions were observed, \sioa{} with the 1.4\,mm receiver and \siob{} with the 3.4\,mm receiver, seen in Figure \ref{sio}. The five molecular outflows we detect are also marked, labeled SIO-a -- e, four of them at both wavelengths and the larger, westernmost outflow, only at 3.4\,mm. In Figure \ref{sio-spectra} we present the spectra of the five outflows, taken toward the center of the outflow for SIO-a, -b and -d, and toward the wings for SIO-c and -e.
From the integrated spectra of the \rxa{} emission within a $4\arcsec\times4''$ region centered at the phase center (Fig. \ref{sio-spectra}f), we determine that the contribution of the ambient gas emission ranges from -43 to -36\,km\,s$^{-1}$. This is also in agreement with the linewidth expected from the ambient gas emission for dense and small cores \citep{garay1999}.

The spatial coincidence of the blue- and redshifted emission of SIO-a with MM1 indicates that it is aligned very close to the line of sight (l.o.s.). This can be clearly seen in the \sioa{} transition at \rxa{}, where the most compact emission is mapped (Fig. \ref{sio}). At this wavelength the red- and blueshifted contours are overlapped, however in the \siob{} transition at \rxb{} only a blueshifted emission peak is seen. There is overlapping redshifted emission, but it does not peak at the same position but rather $\sim$0.5\arcsec{} to the southeast of the blueshifted peak.
The position of the \sioa{} peaks as well as the blueshifted \siob{} peak match the position of MM1, and this is likely the driving source of this outflow.
The spectrum at the center position (Fig. \ref{sio-spectra}a) exhibits a broad emission with a FWHM linewidth of 7\,km\,s$^{-1}$, showing also a strong peak at the systemic velocity and a redshifted narrow (FWHM of 1.3 \,km\,s$^{-1}$) feature peaking at $\sim-32$\,km\,s$^{-1}$.

For SIO-b on the other hand, we detect and resolve a blue- and redshifted peak only at \rxb{} while at \rxa{} we only resolve a blueshifted peak. The peaks at both wavelengths are resolved, and the source MM2 is the most likely driving source. Its central spectrum (Fig. \ref{sio-spectra}b) shows clear blue- and a redshifted peaks. This outflow also seems to be aligned close to the l.o.s., according to the position of the blue- and redshifted emission peaks without any emission at the V$_{\mathrm{LSR}}$.

The outflow SIO-c is detected in both SiO transitions, although at \rxb{} the redshifted lobe is faint. The outflow is aligned in the east--west direction, its axis defined as the straight line joining the blue and red emission peaks at \rxb{}. The source MM4 lies on the outflow axis and between the lobes, thus this may be the driving source. The \sioa{} spectrum from this outflow (Fig. \ref{sio-spectra}c) shows broad blue- and redshifted peaks. The blue peak is broad ($\sim 5.5$\,km\,s$^{-1}$) and asymmetric towards bluer velocities, its maximum at $-44$\,km\,s$^{-1}$, while the redshifted peak is symmetric and narrower, with a FWHM linewidth of 4.8\,km\,s$^{-1}$ peaking at $-36$\,km\,s$^{-1}$.

For SIO-d we do not detect any millimetric source that could be driving it. This outflow is best traced at \rxb{} in the \siob{} transition, where both blue- and redshifted lobes are clearly seen, while at \rxa{} just the redshifted emission is traced. The central spectrum for this outflow (Fig. \ref{sio-spectra}d) shows broad ($\sim 5.5$\,km\,s$^{-1}$) double-peaked emission centered at the systemic velocity.
On larger scales \citet{ridge2001} found a CO outflow spanning 1.73\,pc in the northwest--southeast direction, the same as SIO-d (see Section \ref{disc_outflows}).

SIO-e is even more intriguing because it is the biggest of the five outflows and we cannot identify a driving source for it in the millimetric or the IR. Like SIO-d it lies in the northwest--southeast direction, matching the large-scale CO emission. Its \siob{} spectrum shows broad blueshifted emission, peaking at $-40.4$\,km\,s$^{-1}$ with a FWHM of 5.6\,km\,s$^{-1}$. The redshifted peak is narrower with a FWHM of 3.4\,km\,s$^{-1}$ and with its maximum at $-38$\,km\,s$^{-1}$. Its driving source should be located relatively far from the two main sources MM1 and MM2, at $\sim$10$^4$ AU from them.

\subsection{Sulfur Dioxide emission}

Figure \ref{so2} panels (a) and (c) show the integrated Sulfur Dioxide (\so{}) emission detected towards \mia{} at \rxa{} and \rxb{} respectively, while in panels (b) and (d) are the velocity maps for each transition. The \so{} emission envelops the whole region where MM1, MM2, MM3 and MM4 are detected, which can be clearly seen in Figure \ref{so2}c. We see a clear velocity gradient in the southeast--northwest direction, indicating that the envelope of \mia{} rotates as a whole and is likely gravitationally bound. This is more noticeable in the \sob{} velocity map (Fig. \ref{so2}d), where the velocity gradient goes smoothly from $-42$ to $-37$\,km\,s$^{-1}$.

The peaks of integrated \so{} emission at \rxa{} (Fig. \ref{so2}a) lie close to the sources MM1, MM2 and MM4. As \so{} is also an indicator of shocked material \citep[e.g.][]{helmich1994} this strengthens the argument for the detection of the outflows traced by SiO, and supports the statement in Section \ref{sio2} that those sources drive the outflows SIO-a, SIO-b and SIO-c, respectively.

The \soa{} velocity map (Fig. \ref{so2}b) shows a strong jump of $\sim4$\,km\,s$^{-1}$ in a thin spatial region southeast of the two main millimetric peaks. The velocity shift is from $\sim -35$\,km\,s$^{-1}$ to $\sim\,-39$\,km\,s$^{-1}$ --the systemic velocity-- in a narrow strip of $\sim \, 1.1\arcsec{} \times 0.2$\arcsec{}. The region north of the jump is where most of the gas is blueshifted while the region south of the jump is completely redshifted. This ``velocity jump'' (see Sec. \ref{velo}) may be the signature of two molecular flows moving in opposite direction, ramming each other and compressing the gas in the region of the jump \citep{klessen2005,heitsch2005,heitsch2006,peretto2006,peretto2007}.

\section{Discussion}
\subsection{Continuum Sources}
\label{cont_disc}

With $\mathrm{L}=2\times10^5 \mathrm{L}_{\odot}$, \mia{} is a high-mass star-forming region, but the (proto)stellar content in its core is unknown because of high obscuration. Since it is a hypercompact \hii{} region, it already has a large (proto)stellar component. While it is not possible to directly measure the masses of the (proto)stars with millimeter data, it is possible to derive the masses of the circumstellar structure.  The low masses that we calculate from the \rxa{} data can be attributed to a circumstellar structure, while the surrounding envelope likely contributes to the larger masses derived from the \rxb{} data. However, we must take into account the ``short-spacing problem'' when calculating the masses of the continuum sources. This caveat, inherent to interferometers, means that a high percentage of the flux is filtered out and lost.  This effect is more severe in the extended configurations (long baselines), as in our case.

Single-dish submillimetric measurements for \mia{} at $850\mu$m with SCUBA \citep{moore2007} indicate a peak intensity $\mathrm{I}_{\nu}\sim9.6$\,Jy\,beam$^{-1}$ with a beam of 14\arcsec{}. Considering a dependence $\mathrm{I}_{\nu}\propto\nu^4$ for the peak intensity, that value corresponds then to $\mathrm{I}_{\nu}\sim1.4$\,Jy\,beam$^{-1}$ at \rxa{} and $\mathrm{I}_{\nu}\sim36$\,mJy\,beam$^{-1}$ at \rxb{}. In our \rxa{} continuum map the flux density recovered within the SCUBA beam is $\mathrm{S}_{\nu}\sim0.12$\,Jy taking the average contribution of the free-free emission into account (see Table \ref{ffree}), meaning that $\sim91\%$ of the flux is lost. Therefore the calculated masses at \rxa{} are actually a lower limit for the current mass of each source.
On the other hand, at \rxb{} we recover $\mathrm{S}_{\nu}\sim36$\,mJy, so with our calibration accuracy, we do not miss too much of the flux within the SCUBA beam. However, within the PdBI main beam of $\sim55$\arcsec{} at \rxb{} we only recover $\sim73$\,mJy, while \citet{moore2007} reports an integrated flux $\mathrm{S}_{\nu}\sim200$\,Jy, implying that although we are recovering most of the central emission, we are filtering out the more extended emission. Nevertheless, at \rxb{} the obtained values for the masses are somewhat more representative of the mass of their envelopes.

The \so{} velocity signature across the core is indicative of overall rotation of the gas associated with MM1 to MM3. Assuming equilibrium between the gravitational and rotational force at the outer radius of the cloud:

\begin{equation}
M_{rot}=\frac{\delta v^2r}{G}
\end{equation}
\begin{equation}
\Rightarrow M_{rot}[\mathrm{M_\odot}]=1.13\;10^{-3}\times \delta v^2[\mathrm{km\,s^{-1}}] \times r[\mathrm{AU}]
\end{equation}
where $r$ is the cloud radius ($\sim2''=4000\,$AU) and $\delta v$ is half the velocity regime observed in the first-moment map ($\sim2.5$\,km\,s$^{-1}$; Fig. \ref{so2}d).

The gravitationally supported mass implied by this observed rotation is M$_{\mathrm{rot}}\sim30$\,M\solar{}, which is on the same order as the derived masses from the \rxb{} continuum emission. Hence, these sources are likely gravitationally bound and may still be accreting from an envelope not seen in our data, further increasing their masses. Also, from the distribution of the gas it is likely that sources MM1, MM2 and MM3 are sharing a common envelope, of which we are tracing the inner parts at \rxb{}. This envelope would have much more of the mass than in the immediate region surrounding each source, but it is being filtered out by the interferometer.

The different percentage of flux filtered out in each wavelength band can be explained the same way as the big (over a factor $\sim10$) difference between the calculated masses at \rxa{} and \rxb{}. Both wavelengths were observed at the same time with the same interferometer configurations, meaning that although the ground baselines are the same, the uv coverage measured in units of wavelength (k$\lambda$) at \rxb{} is more compact than at \rxa{}, therefore tracing more extended components in the region.

Even if we are recovering most of the compact flux at \rxb{}, we are likely mapping the surrounding envelopes close to the already formed (proto)stellar objects and therefore cannot state the actual masses of the (proto)stellar objects but only give a lower limit of the mass of the gas and dust associated largely with the individual sources. The same can be applied for the column density, however in this case --despite the caveats mentioned-- the values we obtain are more realistic for a region like \mia{}. Our results of $N(H_2)\sim10^{25}$\,cm$^{-2}$ at \rxb{}, are 2--3 magnitudes higher than for low-mass star-forming regions, for which the derived column densities at $\sim$3\,mm are of the order $N(H_2)\sim10^{22-23}$\,cm$^{-2}$ \citep[e.g.: ][]{harjunpaa1996,motte1998,kontinen2000}.  Also our results are about two orders of magnitude higher than the critical column density of 1\,g\,cm$^{-2}$ required to form high-mass stars under Milky Way conditions calculated by \citet{krumholz2008}, which corresponds to $N(H_2)=3\times10^{23}$\,cm$^{-2}$, suggesting that within the sources we are mapping, (proto)stars of over 10\,M\solar{} are forming.

\begin{table}[h]
\renewcommand{\arraystretch}{1.2}  
\begin{minipage}[]{\columnwidth}
\caption{Millimetric, radio, NIR and MIR counterparts.}
\label{table-assoc}
\centering
\begin{tabular}{cccc}
\hline\hline
MM & Q\footnote{From \citet{vandertak2005a}} & NIR\footnote{From \citet{megeath2005}} & MIR$^a$\\
\hline
1 & 5 & 1 & 1 \\
2 & 3 & 2 & 2 \\
3 & $1+2$ & 2a & \ldots \\
4 & \ldots & \ldots & \ldots \\
5 & $1+2+3$ & \ldots & \ldots \\
6 & 4 & 3 & 3\\
\hline\hline
\end{tabular}
\renewcommand{\footnoterule}{} 
\end{minipage}
\end{table}

Radio observations for this region are also available. Comparing the positions of the five radio sources detected with the VLA at 43\,GHz ($Q$-band) by \citet{vandertak2005a} and the positions in Table \ref{table-sources}, we can identify the following: MM1~$\rightarrow$~Q5, MM2~$\rightarrow$~Q3 and MM6~$\rightarrow$~Q4 (see Fig.~\ref{cont}). The source MM3 lies between Q1 and Q2 and is not a compact source but rather stretched out, so it appears to be the joint contribution from Q1 and Q2, which could not be individually resolved with our spatial resolution (see Table \ref{table-assoc}).

At shorter wavelengths, \citet{vandertak2005a} found three MIR sources in \mia{} with the Keck I telescope, while \citet{megeath2005} found seven NIR sources with the Hubble Space Telescope, three of them which match the aforementioned MIR sources.

Because absolute astrometry was not available for these NIR and MIR sources, to compare them with our MM sources we refer to their relative positions, shown in Table \ref{table-astrometry}. The separation between MM1 and MM2 is $1.16'' \pm 0.10''$ at a position angle (P.A.) of $214 \pm 5$ degrees, being the only match for the pair MIR1 and MIR2 with a mean separation of $1.12'' \pm 0.07''$ and a P.A. of $216.8 \pm 1.7$ degrees, and for the pair NIR~1 and NIR~2 with a separation of $1.23''$ and a P.A. of 217 degrees.  Similarly, the pair MM1--MM6 with a separation of $2.65'' \pm 0.10''$ and $\mathrm{P.A.}= 187 \pm 2$ degrees is the only match for the pair MIR1--MIR3 with a mean separation of $2.70'' \pm 0.05''$ and a P.A. of $187.2 \pm 0.7$ degrees and for the pair NIR~1--NIR~3 with a separation of $2.7''$ and $\mathrm{P.A.}= 189$ degrees. Therefore we can identify MIR1 and NIR~1 with MM1, MIR2 and NIR~2 with MM2, and MIR3 and NIR~3 with MM6.
Comparing the separation of $0.94'' \pm 0.10''$ and $\mathrm{P.A.}= 235$ between MM1 and MM3, with the separation of $0.98''$ and $\mathrm{P.A.}= 224$ between NIR~1 and NIR~2a, we can also identify source NIR~2a with MM3. For the rest of the NIR sources we do not find a millimetric counterpart (see Table~\ref{table-assoc}).

The only millimetric source without any NIR, MIR or radio counterpart is MM4, hence it is a new detection. Its mass could not be corrected by the free-free emission contribution and it is approximately half as massive as MM3.

Considering the region encompassed by MM1 and MM2, we estimate a lower limit for the (proto)stellar density. Assuming spherical symmetry, we detect the four sources MM1, MM2, Q1 and Q2 (joined in MM3) within a $1.16''$ or \mbox{$\sim2100$\,AU} diameter region, corresponding to a (proto)stellar density of \mbox{$\sim7\times10^6$\,protostars\,pc$^{-3}$}. Even counting just the three resolved sources MM1, MM2 and MM3 we obtain \mbox{$\sim5\times10^6$\,protostars\,pc$^{-3}$}.
It is not the first time that (proto)stellar densities higher than the typical $10^4$ stars\,pc$^{-3}$ for young clusters \citep{lada2003} are found. \citet{beuther2007d} calculated a density of $\sim1.4\times10^5$ protostars\,pc$^{-3}$ in the system G29.96--0.02.
Although high, this (proto)stellar density is still below the ``critical'' value of \mbox{$\sim10^8$\,protostars\,pc$^{-3}$} predicted by the merging scenario \citep{bonnell1998}. However it is the first time that densities higher than the required \mbox{$\sim10^6$\,protostars\,pc$^{-3}$} to induce high-mass close binary mergers \citep{bonnell2005} are obtained. While the observations of \mia{} show no evidence of the presence of such binary systems, we have shown that such densities can be achieved in high-mass star-forming regions, and with further improvement in the spatial resolution, (proto)stellar densities as high --or even higher-- may be a common phenomenon.

\begin{table}[h]
\renewcommand{\arraystretch}{1.2}  
\begin{minipage}[]{\columnwidth}
\centering
 \caption{Relative astrometry of the known sources in \mia{}.}
\label{table-astrometry}
\begin{tabular}{lcc}
\hline\hline
 & sep & PA \\
Sources & (arcsec) & (deg)\\
\hline
MM1--MM2 & 1.16 & 214 \\
MIR1--MIR2\footnote{Mid-infrared sources from \citet{vandertak2005a}} & 1.12 & 216.8\\
NIR1--NIR2\footnote{Near-infrared sources form \citet{megeath2005}} & 1.23 & 217 \\
\hline
MM1--MM6 & 2.65 & 187 \\
MIR1--MIR3 & 2.70 & 187.2 \\
NIR1--NIR3 & 2.7 & 189 \\
\hline
MM1--MM3 & 0.94 & 235 \\
NIR1--NIR2a & 0.98 & 224 \\
\hline\hline
\end{tabular}
\renewcommand{\footnoterule}{}
\end{minipage}
\end{table}

\mia{} is surrounded by a dense embedded cluster of intermediate- to low-mass stars: \citet{megeath1996} calculated a stellar surface density of $\sim0.25\times10^4$\,pc$^{-2}$ for a cluster radius of $21''$, a quarter of the \mbox{$\sim10^4$\,pc$^{-2}$} density calculated by \citet{megeath2005} from the NIR sources they found with a maximum projected separation of $\sim5,600$\,AU. Again within a \mbox{$\sim2100$\,AU} diameter region, we calculate a (proto)stellar surface density of \mbox{$\sim4\times10^4$\,protostars\,pc$^{-2}$} considering just the three millimetric sources previously mentioned, revealing a density gradient towards the center of the cluster. Assuming that this embedded cluster around \mia{} forms an independent subcluster within the W3-Main region, the fact that there are already low-mass stars in the outskirts of \mia{} while high-mass stars are still forming in its center supports the hypothesis that the low-mass stars form before their high-mass counterparts \citep[e.g.,][]{kumar2006}.

The projected separation between sources ranges from $\sim$750\,AU for MM2--MM3 to \mbox{$\sim$4700\,AU} for MM1--MM6. Compared to the median radius of 40,000\,AU for the farthest outlying member of the 14 trapezia identified by \citet{abt2000}, \mia{} has significantly smaller projected separations than those observed in optically visible trapezia. 
Therefore we have several compact sources sharing a common envelope from which gas may still be accreting in a region with a high (proto)stellar density beyond the point where close binary mergers can be induced, and with column densities higher than the theorized threshold to form high-mass stars. Thus, conditions are favorable for the formation of high-mass stars.

\subsection{Outflows}
\label{disc_outflows}
For the five SiO outflows mapped in this region, we identify the driving sources of just three of them. Outflows SIO-a, SIO-b and SIO-c are driven by sources MM1, MM2 and MM4 respectively, while for outflows SIO-d and SIO-e we do not detect candidates for their driving sources. For SiO-e however, \citet{tieftrunk1998} have detected a NH$_3$~(1,1) emission peak (labeled 1 in their Figures 4 and 5) near the center of this outflow, which may be hosting the driving source of SiO-e.

The determination of the outflow parameters is subject to a big number of errors, including that it is often difficult to separate the outflowing gas from the ambient gas and that the inclination angles of the flows are also often unknown. In this case we also have the caveat that high amounts of flux are filtered out by the interferometer.
To get an estimation of the effect of this we compare our results with the measurements of \citet{gibb2007}. They report an integrated intensity of $2.4$\,K\,km\,s$^{-1}$ for \sioa{}, while we recover only $0.4$\,K\,km\,s$^{-1}$ within our primary beam. Although apparently the integrated intensity given by \citet{gibb2007} covers a slightly larger region, the main emission stems from an area covered by approximately twice our 22\arcsec{} primary beam at 217\,GHz, implying a great fraction of missing flux also in \sioa{}. On the other hand, for \siob{} they measure $2.5$\,K\,km\,s$^{-1}$ but with most of the emission within a region covered by our 55\arcsec{} primary beam at 87\,GHz, in which we measure $1.7$\,K\,km\,s$^{-1}$. Again, we are losing a high amount of flux, although less than in \sioa{}.

We derived the physical parameters for the small-scale outflows detected here. To obtain the H$_2$ column densities we first have to calculate the SiO column densities for each outflow, for which we follow the LTE approach described by \citet{irvine1987}. Since we do not have estimates of the excitation temperature, we assume a usual value for the outflows of 30\,K for our calculations \citep{beuther2002b}, resulting in the formula:
\begin{equation}
	N_{\mathrm{SiO(2-1)}}=1.3\times10^{12}\int{T\mathrm{d}v}
\end{equation}
where $\int{T\mathrm{d}v}$ is the \siob{} integrated intensity in K\,km\,s$^{-1}$. We used the \siob{} data because it has recovered the most flux. The obtained results are shown in Table \ref{table-energy}. Adopting a SiO abundance ratio to obtain the H$_2$ column density is not straightforward. The SiO abundance has a big uncertainty, with reported values ranging from SiO/H$_2\sim10^{-7}$ \citep{mikami1992,zhang1995} for shocked, energetic regions to SiO/H$_2\sim10^{-12}$ \citep{ziurys1987,ziurys1989} in quiescent, cold dark clouds. To take into account such uncertainty, our calculations were done assuming SiO/H$_2=10^{-7}$ and SiO/H$_2=10^{-9}$, a range suitable for a high-luminosity region such as \mia{}. Although the values we obtain are too low for a region like \mia{} with a total luminosity of $\sim10^5\mathrm{L}$\solar{}  and are more typical in regions with lower luminosity \citep[see e.g. ][]{ridge2001,gibb2007,palau2007,beltran2008}, we have to keep in mind that aside from the caveats mentioned above, we are tracing smaller scales opposed to large-scale outflow.

In general, our parameters are several orders of magnitude lower than for the CO outflow detected by \citet{ridge2001} even assuming extremely low SiO abundances. 
From the CO measurements, the outflow has a mechanical energy \mbox{E$_{\mathrm{CO}}\sim10^{49}$\,erg}, while for a SiO abundance of $10^{-9}$, adding the contribution of all the outflows we obtain that \mbox{E$_{\mathrm{SiO}}/$E$_{\mathrm{CO}}\sim4\times10^{-4}$}. From the values in Table \ref{table-energy} it is evident that the rest of the parameters follow the same trend. Only at unrealistically low SiO abundances are our results comparable with the CO outflow.

\begin{table}[h]
\renewcommand{\arraystretch}{1.2}
 \begin{minipage}{\columnwidth}
\centering
\caption{Properties of the small-scale outflow features}
\label{table-energy}
\begin{tiny}
\begin{tabular}{lccccc}
\hline\hline
Outflow & M$_{out}$ & E & $\dot{M}_{out}$ & F$_{m}$ & L$_{m}$ \\
 & {\tiny (M$_{\odot}$)} & {\tiny (erg)} & {\tiny (M$_{\odot}$\,yr$^{-1}$)} & {\tiny (M$_{\odot}$\,yr$^{-1}$\,km\,s$^{-1}$)} & {\tiny (L$_{\odot}$)} \\
\hline \hline
\multicolumn{6}{c}{SiO/H$_2=10^{-7}$}\\
\hline
SIO-a + SIO-b\footnote{Treated jointly because they cannot be differentiated in the redshifted \siob{} emission. Also, because they are aligned close to the l.o.s., the parameters for which the outflow size is needed were calculated assuming an average size based on the other three outflows.} & 0.04 & 4(43) & 4(-5) & 2(-4) & 0.25 \\
SIO-c & 0.03 & 5(43) & 5(-5) & 7(-4) & 0.80 \\
SIO-d & 0.03 & 1(43) & 3(-5) & 2(-4) & 0.09 \\
SIO-e & 0.08 & 3(43) & 1(-5) & 6(-5) & 0.03 \\
\hline
\multicolumn{6}{c}{SiO/H$_2=10^{-9}$}\\
\hline
SIO-a + SIO-b$^a$ & 4 & 4(45) & 4(-3) & 2(-2) & 25 \\
SIO-c & 3 & 5(45) & 5(-3) & 7(-2) & 80 \\
SIO-d & 3 & 1(45) & 3(-3) & 2(-2) & 9 \\
SIO-e & 8 & 3(45) & 1(-3) & 6(-3) & 3 \\
\hline\hline
\end{tabular}
\end{tiny}
\renewcommand{\footnoterule}{} 
\end{minipage}
\end{table}

The SiO outflow that \citet{gibb2007} detect is aligned in the east--west direction, and the CO outflow also has an east--west component peaking at $\sim$35$''$ east of the main source extending to the west, as well as a NW--SE component. In Fig. \ref{sio} we can see that the \siob{} emission follows a similar distribution, with east--west (outflow SIO-c) and NW--SE (outflows SIO-d and -e) components. As mentioned before we do not detect driving sources for SIO-d and -e. Because of their orientation these two outflows could be delineating the high-density regions of the large-scale CO outflow and although we cannot exclude this option, we believe that they are two separate features. Furthermore, \citet{tieftrunk1998} detects an ammonia signature close to the projected center of SIO-e which may be hosting its driving source. With our data we cannot confidently point to any of our detected dust sources as the driving source of the large-scale CO outflow and it is possible that this CO outflow is the result of the mix of all the small-scale outflows and therefore might have driving sources opposed to a single driving source.

As shown in Section \ref{sio2} the outflows SIO-a and SIO-b are aligned very close to the l.o.s.. This situation explains the fact that despite the high column densities toward MM1 and MM2 (see Table \ref{table-sources}), which correspond to visual extinctions of several thousand magnitudes, NIR and MIR counterparts are detected. The cavities carved out by the outflows have reduced the extinction along the l.o.s., allowing us to see hot dust in the interior of the protostellar envelopes in the NIR and MIR, either directly or through scattered light.

In the case of SiO-b, the outflow also would explain the occurrence and location of NIR 2b, below NIR 2/MM2. That near-infrared source can be interpreted as scattered light from the outflow cavities. Similarly in the case of source NIR 2a, it is close to the proper motion radio source IRS 5a detected by \citet{wilson2003} with the VLA. They propose that IRS 5a/NIR 2a was formed by or in an outflow originating south of its position, which agrees with the localization of our outflow SiO-b.

From the H$_2$O masers they detect, \citet{imai2000} suggest two outflows in the north--south direction in this region. However from our data we cannot identify counterparts for the position of the H$_2$O masers in this region, nor outflows in a north--south direction.

\subsection{Velocity jump}
\label{velo}
Figure \ref{so2-jump} shows a postion-velocity diagram across the velocity jump seen in \soa{}. Two distinct velocity components are seen, one peaking at $\sim-35$\,km\,s$^{-1}$ and the other peaking at $\sim-39$\,km\,s$^{-1}$, spatially located at both sides of the velocity jump, denoted by the black dashed line. The velocity jump has also been confirmed at higher frequencies, in SO$_2$\,(8--7) and HCN\,(4--3) observations taken with the SMA towards \mia{} (T. Bourke, priv. comm.), and could be explained by the theory of gravoturbulent star formation \citep[and references therein]{heitsch2005,heitsch2006,klessen2005,maclow2004}, which states that protostellar cores might form by turbulent ram pressure compression in the regions behind the fronts of converging turbulent flows.

\begin{figure}[h]
	\centering
	\includegraphics[angle=-90, width=\columnwidth]{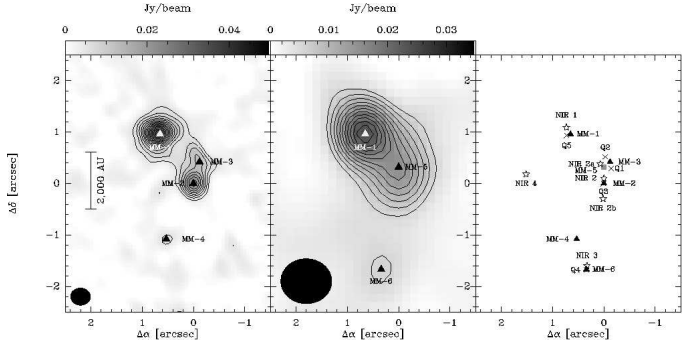}
	\caption{\underline{Left and center panels}: Continuum maps of \mia{} at \rxa{} and \rxb{} respectively. Contour levels for the \rxa{} continuum start at $4\sigma$ increasing by $2\sigma$ steps, and at $3\sigma$ increasing by $1\sigma$ steps for the \rxb{} continuum. Beams are shown at the bottom left of each panel. Filled triangles mark the millimetric sources detected (see Table \ref{table-sources}). \underline{Right panel}: Relative positions of the different sources in \mia{}. Filled triangles are the millimetric sources detected, stars are the NIR sources detected by \citet{megeath2005} and crosses are the radio sources detected by \citet{vandertak2005a}. The grey square marks source MM5, the joint contribution of sources MM2 and MM3. In Table \ref{table-assoc} are the identified counterparts, discussed in the text.}
	\label{cont}
\end{figure}

\begin{figure}[h]
	\centering
	\includegraphics[angle=-90, width=\columnwidth]{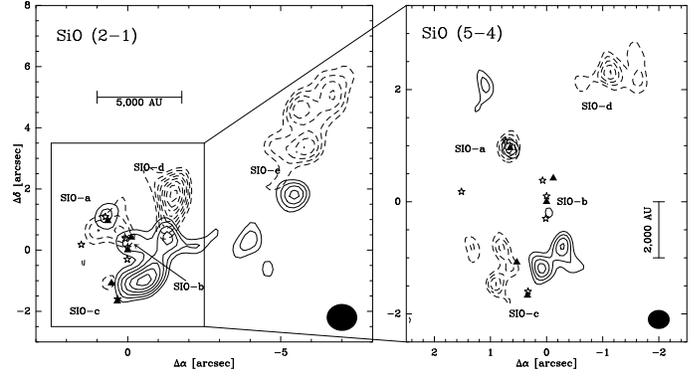}
	\caption{Silicon Monoxide emission detected towards \mia{}. The left panel shows the \rxb{} \siob{} transition and the right panel shows the \rxa{} \sioa{} transition. Solid contours outline the blueshifted emission and dashed contours the redshifted emission. The blueshifted emission ranges from $-58$ to $-43$\,km\,s$^{-1}$ and the redshifted emission ranges from $-36$ to $-28$ km\,s$^{-1}$. In the left panel contouring starts at the 4$\sigma$ level and in the right panel at the 3$\sigma$ level, increasing in 1$\sigma$ steps for both. The beams are represented at the bottom right corner of each panel, respectively. Labeled SIO-a through -e are the five outflows detected, filled triangles are the millimetric sources shown in this paper and the stars mark the NIR sources from \citet{megeath2005}.}
	\label{sio}
\end{figure}

\begin{figure}[h]
	\centering
	\includegraphics[angle=-90, width=0.45\columnwidth]{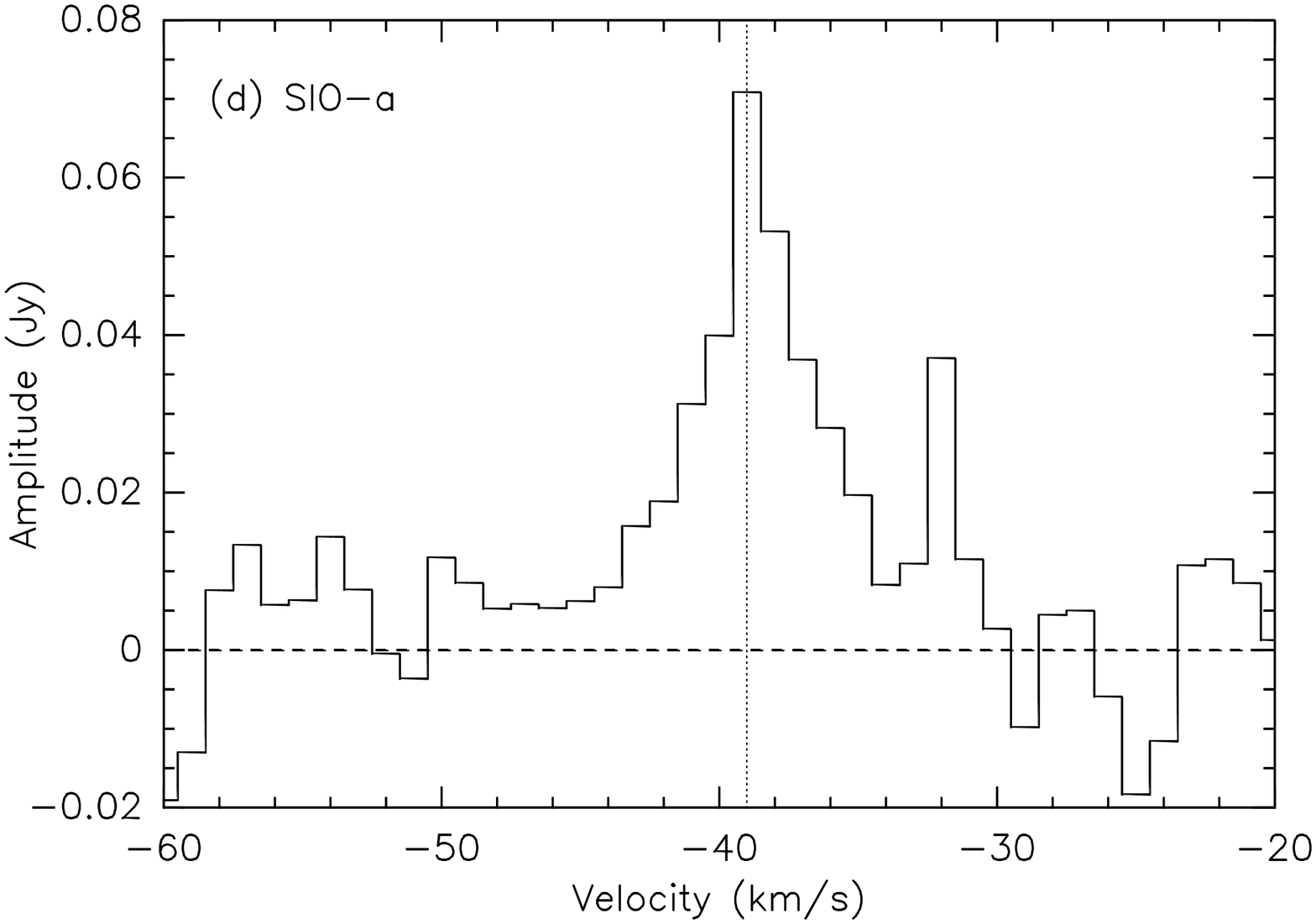}
	\includegraphics[angle=-90, width=0.45\columnwidth]{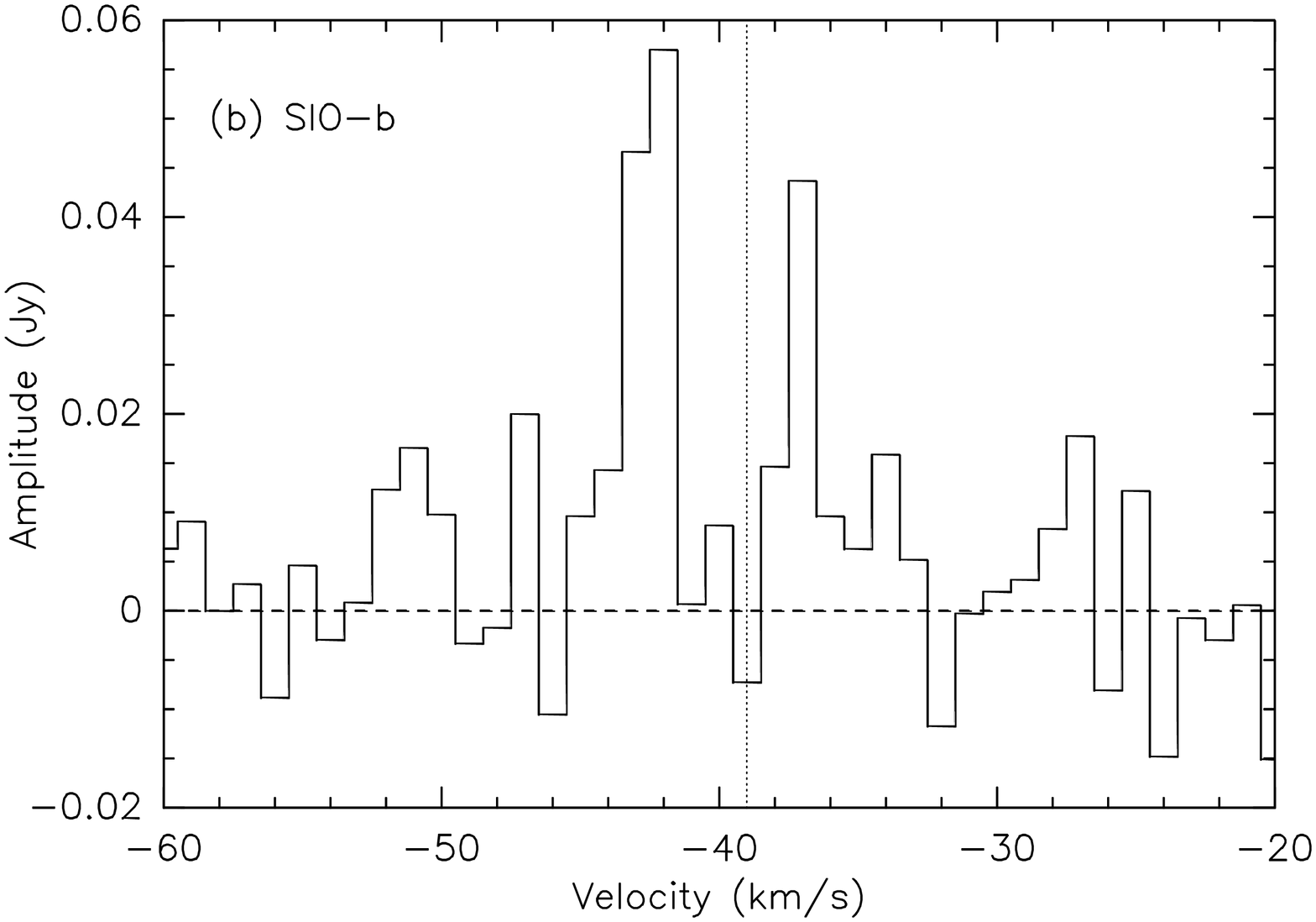}\\
	\vspace{0.3cm}
	\includegraphics[angle=-90, width=0.45\columnwidth]{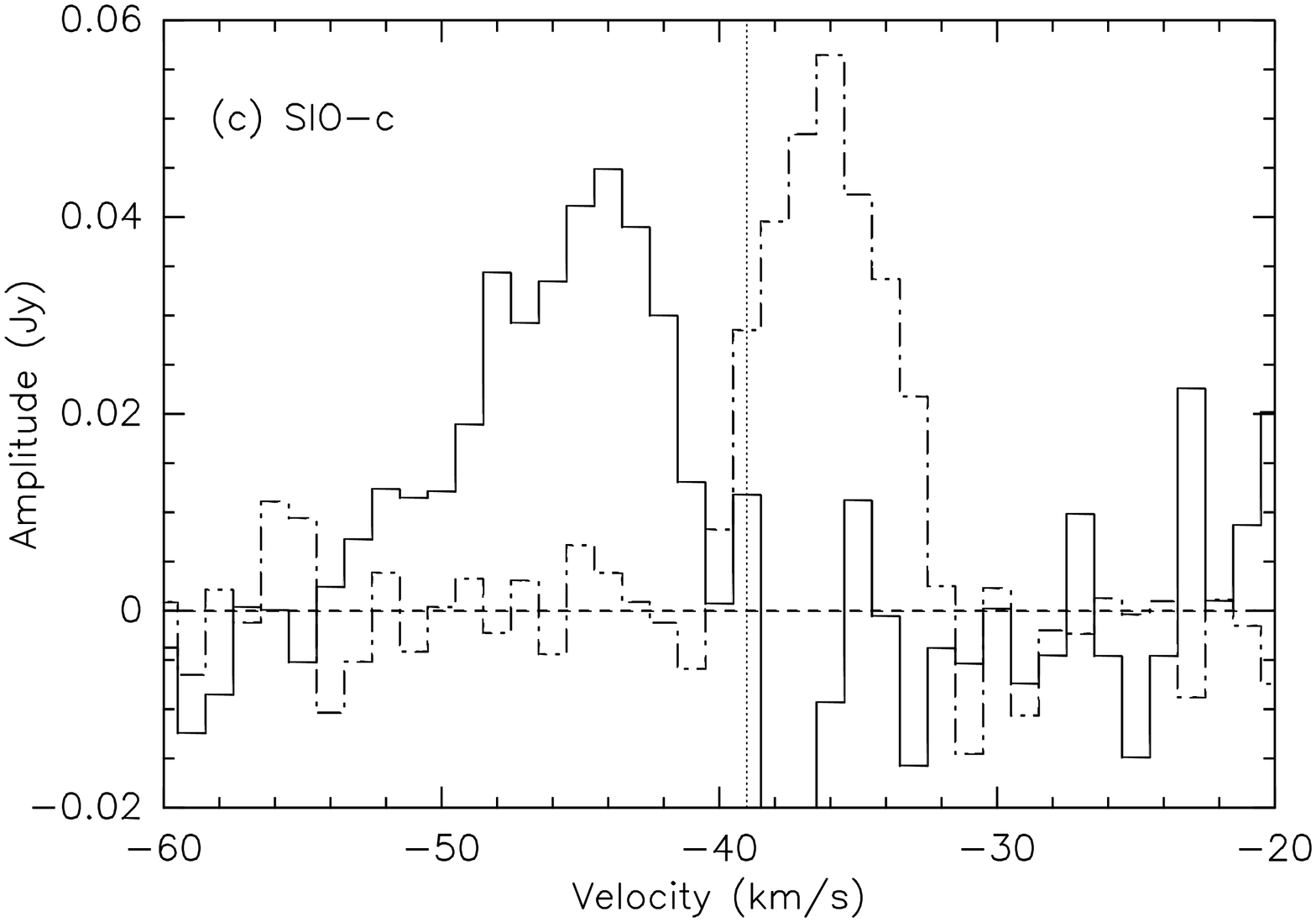}
	\includegraphics[angle=-90, width=0.45\columnwidth]{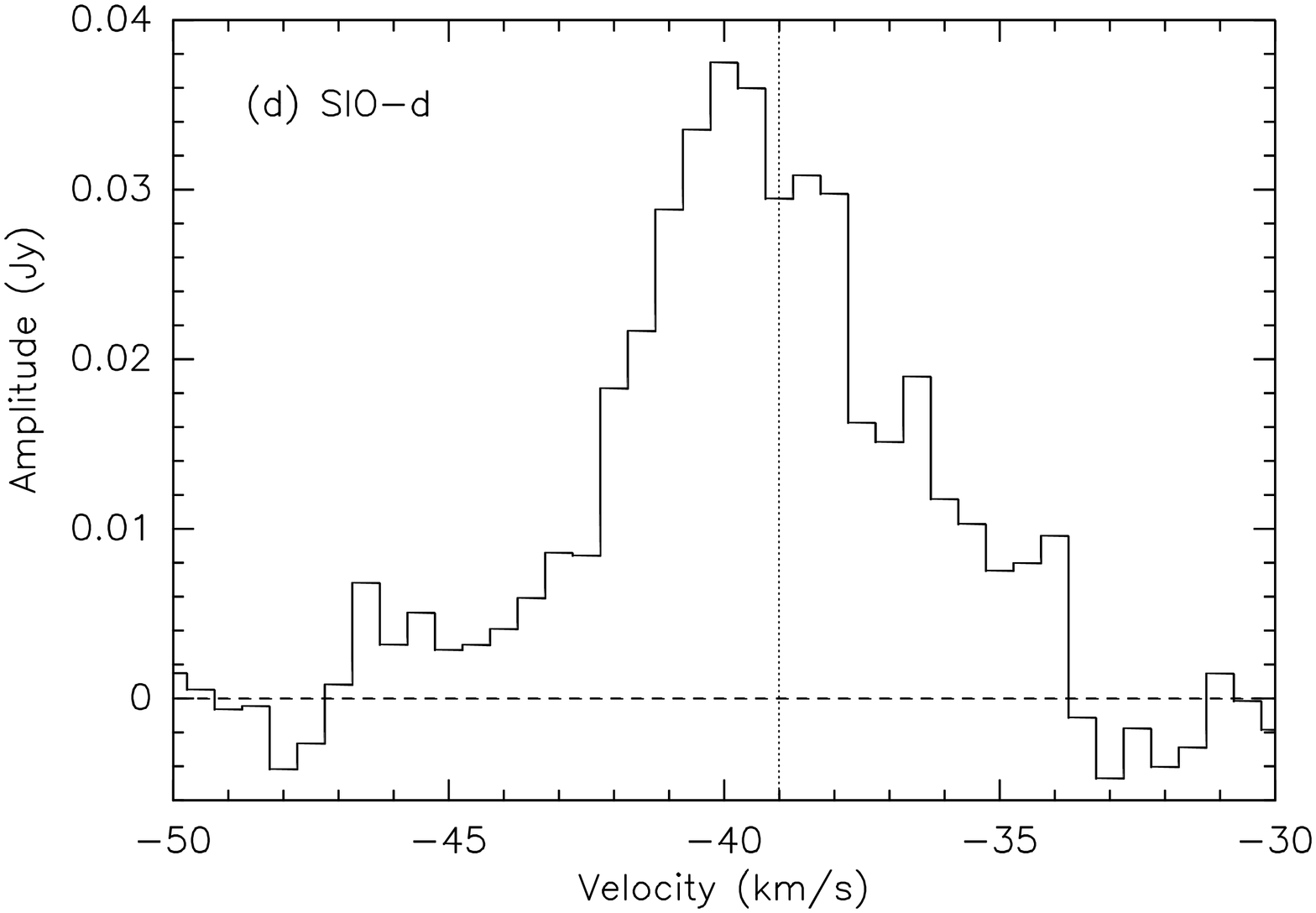}\\
	\vspace{0.3cm}
	\includegraphics[angle=-90, width=0.45\columnwidth]{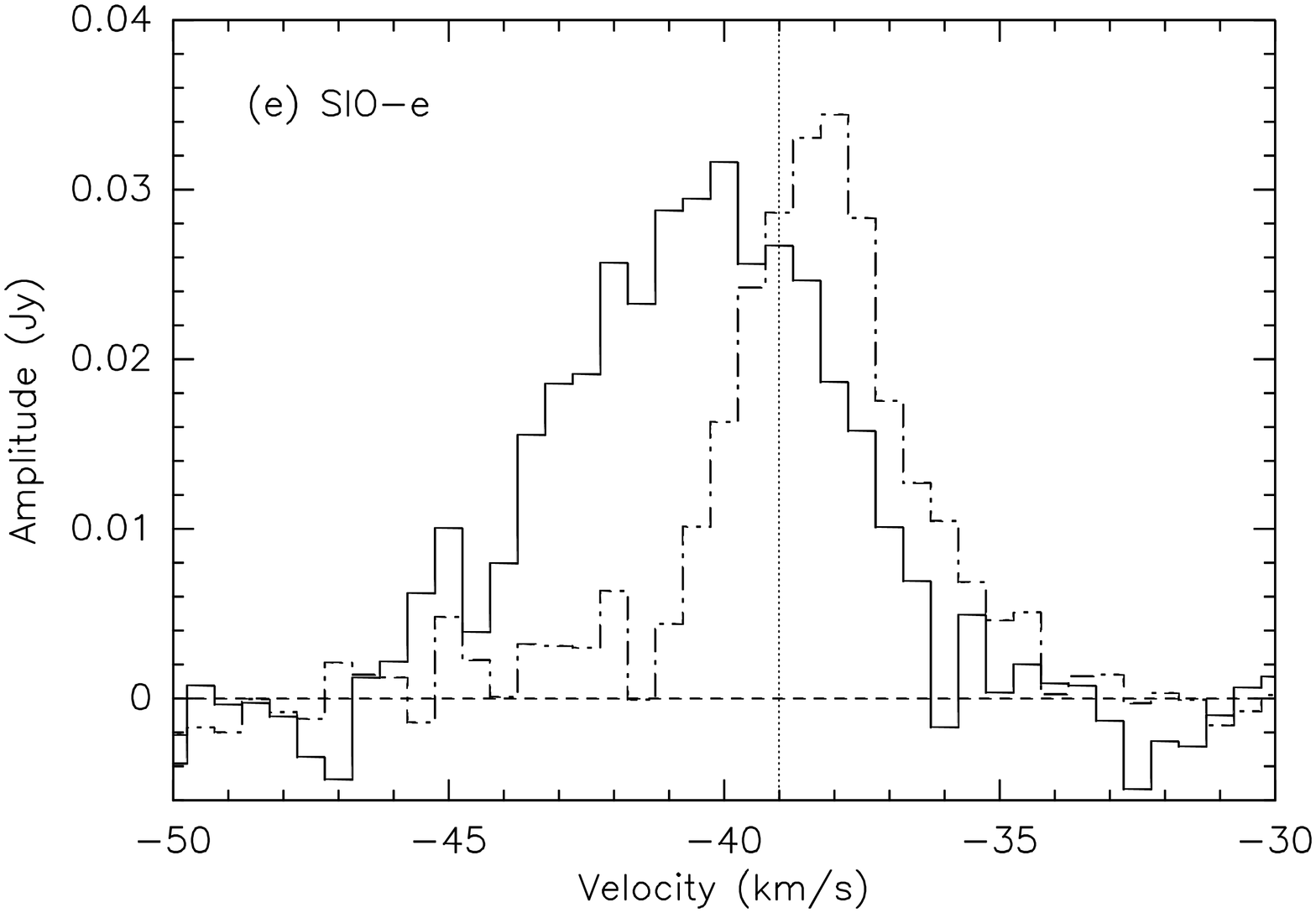}
	\includegraphics[angle=-90, width=0.45\columnwidth]{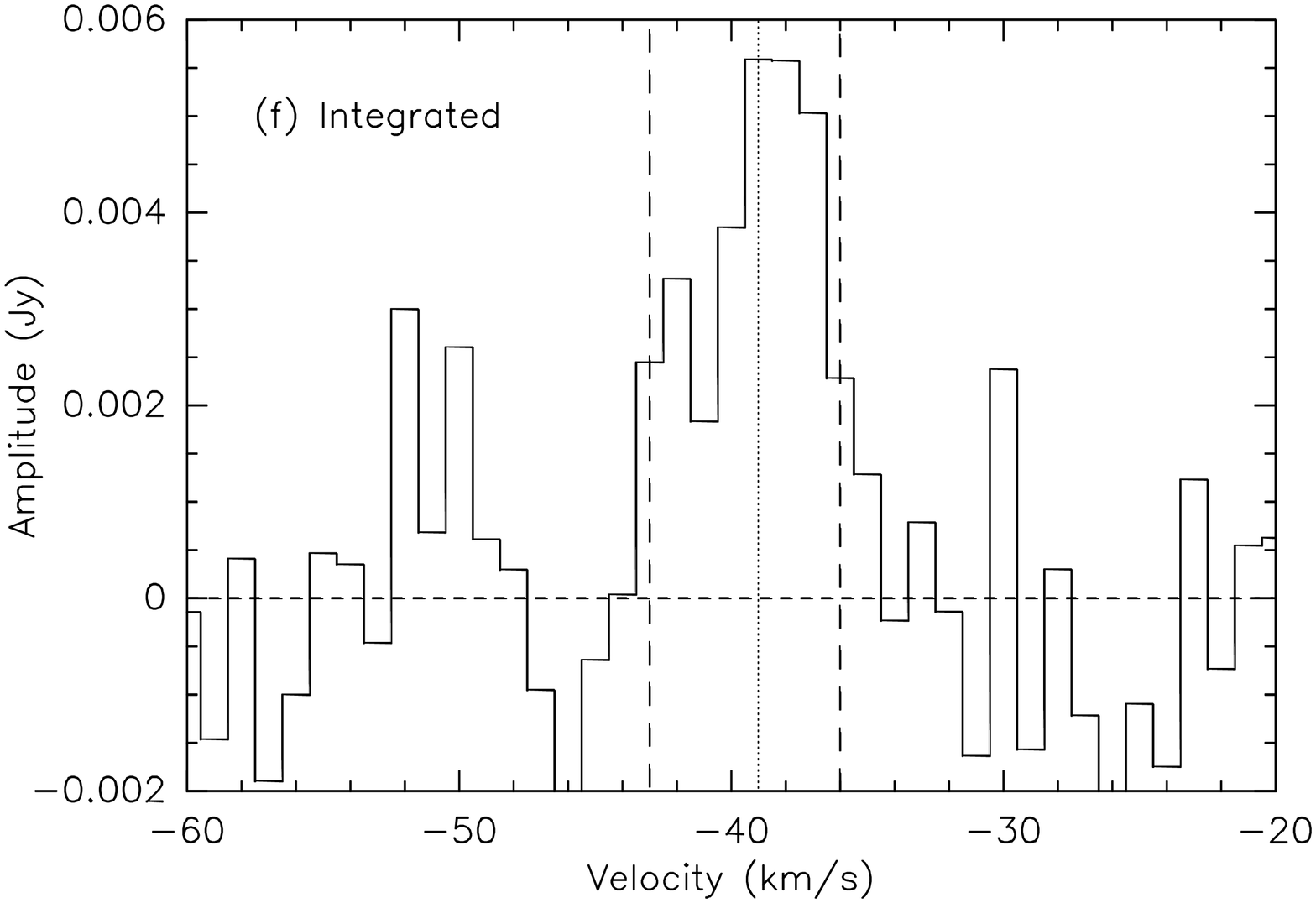}
	\caption{SiO spectra of the five detected outflows. In panels (a), (b) and (c) are the \sioa{} spectra of outflows SIO-a, -b and -c respectively. In panels (d) and (e) are the \siob{} spectra of the outflows SIO-d and e respectively and in panel (f) is the spectrum of the \sioa{} integrated emission in a $4''\times4''$ region centered at the phase center. The vertical dotted line in each panel marks the systemic velocity of $-39$\,km\,s$^{-1}$. In panels (c) and (e) the solid line is the blueshifted emission and the dot-dashed line is the redshifted emission. In panel (f) the long-dashed lines demark the velocity regime for the ambient gas.  The spectra were taken toward the center of the outflow for SIO-a, -b and -d, and toward the wings for SIO-c and -e.}
	\label{sio-spectra}
\end{figure}

\begin{figure}[h]
	\centering
	\includegraphics[angle=-90, width=\columnwidth]{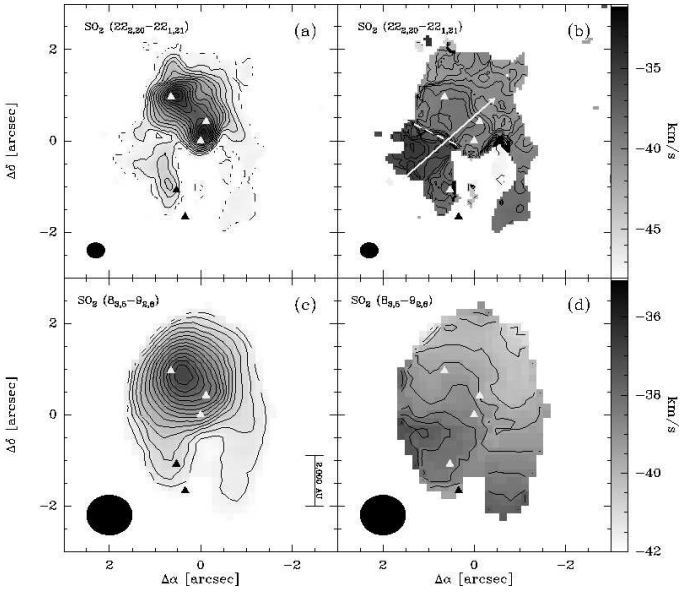}
	\caption{Sulfur Dioxide emission detected towards \mia{}. At \rxa{} is the SO$_2$(22$_{2,20}$--22$_{1,21}$) transition (upper half) and at \rxb{} is the SO$_2$(8$_{3,5}$--9$_{2,8}$) transition (lower half). Panels (a) and (c) show the integrated emission, with contour steps of $0.2$\,Jy\,beam$^{-1}$\,km\,s$^{-1}$ for (a) and $0.1$\,Jy\,beam$^{-1}$\,km\,s$^{-1}$ for (c). In panels (b) and (d) is the velocity distribution of the emission, from -48 to -33\,km\,s$^{-1}$ for (b) and from -42 to -37\,km\,s$^{-1}$ for (d), in both panels the contour step is 0.5\,km\,s$^{-1}$. The respective beams are shown at the bottom-right corner of each panel, respectively. Filled black and white triangles are the mm sources from Table \ref{table-sources}. The dashed white line in panel (b) marks the velocity jump discussed in the text.}
	\label{so2}
\end{figure}

\begin{figure}[h]
	\centering
	\includegraphics[angle=-90, width=\columnwidth]{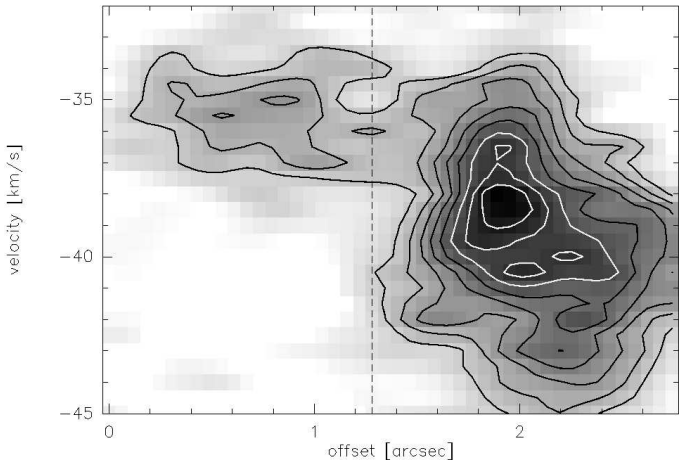}
	\caption{\soa{} position velocity diagram along the line shown in Fig. \ref{so2}(b) with a white arrow. The velocity jump described in the text can be seen here, where an emission feature is peaking at $\sim-35$\,km\,s$^{-1}$ while another feature is peaking at $\sim-39$\,km\,s$^{-1}$, spatially at either side of the jump, denoted by the black dashed line. Contours start at 20\% of the maximum flux and continue in 10\% steps.}
	\label{so2-jump}
\end{figure}

From Figures \ref{cont} and \ref{so2}b we can see that the main sources of \mia{} are located next to the region where the velocity jump occurs, distributed roughly parallel to the jump. This is what would be expected if their collapse were triggered by the compression of converging flows.
Furthermore, numerical models \citep{klessen2005} predict that one of the observational signatures of such compression by ram pressure is localized maxima of the l.o.s. velocity dispersion in the low column density gas in the outskirts of the core. According to the continuum maps (Fig. \ref{cont}), the region where this velocity jump occurs is in the outskirts of the cores, where the dust emission and the gas column density are low.
This suggests that we may be looking at two molecular flows converging and triggering the star formation in \mia{}.

At larger spatial scales, a similar signature was found by \citet{peretto2006,peretto2007} in their study of NGC 2264--C at millimetric wavelengths, where they found in N$_2$H$^+$ and H$^{13}$CO$^+$ emission a velocity jump of $\sim2$\,km\,s$^{-1}$ over a very narrow spatial region. They modeled this jump as the signature of the large-scale, axial collapse of NGC 2264--C along its long axis and towards its center, with an inclination angle of $\sim45^{\circ}$. It would also trace a possible dynamical interaction between protostellar sources at its center, where they detected a millimeter continuum peak. The case of \mia{} could be similar but observed with higher spatial resolution. Making an analogy between \citet{peretto2007} and \mia{}, sources MM1, MM2 and MM3 would correspond to their sources CMM3 and CMM13, and we would just be mapping the region of the jump from redshifted emission to the systemic velocity, which corresponds in their case to the transition from CMM4 to CMM13--CMM3.

\section{Summary and Conclusions}
In the continuum maps of \mia{} obtained with the PdBI at \rxa{} and \rxb{} we detect 5 individual sources.
Their calculated absolute masses are strongly affected by the spatial filtering inherent to the interferometric technique, however from the relative separations of the three strongest sources we propose the scenario of a Trapezium-like system, with at least 3 objects enclosed within a $\sim$2000\,AU radius, with a high (proto)stellar density of \mbox{$\sim5\times10^6$\,protostars\,pc$^{-3}$}, and sharing a common envelope from which they may still be accreting gas.

Four of the five millimetric sources are identified as counterparts of sources previously detected in NIR, MIR and radio wavelengths by \citet{megeath2005} and \citet{vandertak2005a}. The remaining fifth source (labeled MM4) is a new detection.

Thanks to the spatial resolution of $\sim$0.36$''$ we can for the first time trace the small-scale flow, detecting five molecular outflows, identifying the driving source of three of them. Although we find on short spatial scales the same alignment and axes as the outflows on large spatial scales previously detected in SiO and CO, within our accuracy and the big uncertainties introduced, we do not find strong evidence to support that we are tracing the inner and denser regions of those outflows, but we cannot rule out that possibility. However these small-scale outflows appear to be different features.

Two of the SiO outflows are close to the l.o.s. direction and each one is being driven by each of the stronger millimetric sources detected. This configuration would explain why these two objects are detected from NIR to radio wavelengths because cavities being carved out by outflows nearly along the l.o.s. have lower extinction, allowing the detection in the NIR and MIR of the hot dust near the collapsing protostar, while the long wavelength emission traces the dusty envelope that still surrounds it.

The \so{} emission and velocity structure indicate that \mia{} still has a gravitationally bound envelope. Also in \so{} we detect a velocity jump of several km\,s$^{-1}$ in a very narrow region adjacent to the continuum sources. This could be the signature of two converging molecular flows, compressing the gas and triggering the star formation, as predicted by the theory of gravoturbulent star formation.
 
\begin{acknowledgements}
      J.A.R. \& H.B. acknowledge support by the \emph{Deut\-sche For\-schungs\-ge\-mein\-schaft, DFG\/} project number BE~2578. J.A.R. also acknowledge support from the \emph{International Max-Planck Research School for Astronomy and Cosmic Physics} at the University of Heidelberg. The authors would like to thank Phillipe Salome of IRAM Grenoble for his help in the first stage of the data reduction.
\end{acknowledgements}

\def\aj{AJ}%
\def\araa{ARA\&A}%
\def\apj{ApJ}%
\def\apjl{ApJ}%
\def\apjs{ApJS}%
\def\ao{Appl.~Opt.}%
\def\apss{Ap\&SS}%
\def\aap{A\&A}%
\def\aapr{A\&A~Rev.}%
\def\aaps{A\&AS}%
\def\azh{AZh}%
\def\baas{BAAS}%
\def\jrasc{JRASC}%
\def\memras{MmRAS}%
\def\mnras{MNRAS}%
\def\pra{Phys.~Rev.~A}%
\def\prb{Phys.~Rev.~B}%
\def\prc{Phys.~Rev.~C}%
\def\prd{Phys.~Rev.~D}%
\def\pre{Phys.~Rev.~E}%
\def\prl{Phys.~Rev.~Lett.}%
\def\pasp{PASP}%
\def\pasj{PASJ}%
\def\qjras{QJRAS}%
\def\skytel{S\&T}%
\def\solphys{Sol.~Phys.}%
\def\sovast{Soviet~Ast.}%
\def\ssr{Space~Sci.~Rev.}%
\def\zap{ZAp}%
\def\nat{Nature}%
\def\iaucirc{IAU~Circ.}%
\def\aplett{Astrophys.~Lett.}%
\def\apspr{Astrophys.~Space~Phys.~Res.}%
\def\bain{Bull.~Astron.~Inst.~Netherlands}%
\def\fcp{Fund.~Cosmic~Phys.}%
\def\gca{Geochim.~Cosmochim.~Acta}%
\def\grl{Geophys.~Res.~Lett.}%
\def\jcp{J.~Chem.~Phys.}%
\def\jgr{J.~Geophys.~Res.}%
\def\jqsrt{J.~Quant.~Spec.~Radiat.~Transf.}%
\def\memsai{Mem.~Soc.~Astron.~Italiana}%
\def\nphysa{Nucl.~Phys.~A}%
\def\physrep{Phys.~Rep.}%
\def\physscr{Phys.~Scr}%
\def\planss{Planet.~Space~Sci.}%
\def\procspie{Proc.~SPIE}%
\let\astap=\aap
\let\apjlett=\apjl
\let\apjsupp=\apjs
\let\applopt=\ao

\bibliographystyle{aa}

\begin{thebibliography}{56}
\expandafter\ifx\csname natexlab\endcsname\relax\def\natexlab#1{#1}\fi

\bibitem[{{Abt} \& {Corbally}(2000)}]{abt2000}
{Abt}, H.~A. \& {Corbally}, C.~J. 2000, \apj, 541, 841

\bibitem[{{Beltr{\'a}n} {et~al.}(2008){Beltr{\'a}n}, {Estalella}, {Girart},
  {Ho}, \& {Anglada}}]{beltran2008}
{Beltr{\'a}n}, M.~T., {Estalella}, R., {Girart}, J.~M., {Ho}, P.~T.~P., \&
  {Anglada}, G. 2008, \aap, 481, 93

\bibitem[{{Beuther} {et~al.}(2002{\natexlab{a}}){Beuther}, {Schilke}, {Menten},
  {Motte}, {Sridharan}, \& {Wyrowski}}]{beuther2002a}
{Beuther}, H., {Schilke}, P., {Menten}, K.~M., {et~al.} 2002{\natexlab{a}},
  \apj, 566, 945

\bibitem[{{Beuther} {et~al.}(2005){Beuther}, {Schilke}, {Menten}, {Motte},
  {Sridharan}, \& {Wyrowski}}]{beuther2002erratum}
{Beuther}, H., {Schilke}, P., {Menten}, K.~M., {et~al.} 2005, \apj, 633, 535

\bibitem[{{Beuther} {et~al.}(2002{\natexlab{b}}){Beuther}, {Schilke},
  {Sridharan}, {Menten}, {Walmsley}, \& {Wyrowski}}]{beuther2002b}
{Beuther}, H., {Schilke}, P., {Sridharan}, T.~K., {et~al.} 2002{\natexlab{b}},
  \aap, 383, 892

\bibitem[{{Beuther} {et~al.}(2007){Beuther}, {Zhang}, {Bergin}, {Sridharan},
  {Hunter}, \& {Leurini}}]{beuther2007d}
{Beuther}, H., {Zhang}, Q., {Bergin}, E.~A., {et~al.} 2007, \aap, 468, 1045

\bibitem[{{Bonnell} \& {Bate}(2005)}]{bonnell2005}
{Bonnell}, I.~A. \& {Bate}, M.~R. 2005, \mnras, 362, 915

\bibitem[{{Bonnell} {et~al.}(1998){Bonnell}, {Bate}, \&
  {Zinnecker}}]{bonnell1998}
{Bonnell}, I.~A., {Bate}, M.~R., \& {Zinnecker}, H. 1998, \mnras, 298, 93

\bibitem[{{Campbell} {et~al.}(1995){Campbell}, {Butner}, {Harvey}, {Evans},
  {Campbell}, \& {Sabbey}}]{campbell1995}
{Campbell}, M.~F., {Butner}, H.~M., {Harvey}, P.~M., {et~al.} 1995, \apj, 454,
  831

\bibitem[{{Feigelson} \& {Townsley}(2008)}]{feigelson2008}
{Feigelson}, E.~D. \& {Townsley}, L.~K. 2008, \apj, 673, 354

\bibitem[{{Frerking} {et~al.}(1982){Frerking}, {Langer}, \&
  {Wilson}}]{frerking1982}
{Frerking}, M.~A., {Langer}, W.~D., \& {Wilson}, R.~W. 1982, \apj, 262, 590

\bibitem[{{Garay} \& {Lizano}(1999)}]{garay1999}
{Garay}, G. \& {Lizano}, S. 1999, \pasp, 111, 1049

\bibitem[{{Gibb} {et~al.}(2007){Gibb}, {Davis}, \& {Moore}}]{gibb2007}
{Gibb}, A.~G., {Davis}, C.~J., \& {Moore}, T.~J.~T. 2007, \mnras, 382, 1213

\bibitem[{{Harjunpaeae} \& {Mattila}(1996)}]{harjunpaa1996}
{Harjunpaeae}, P. \& {Mattila}, K. 1996, \aap, 305, 920

\bibitem[{{Heitsch} {et~al.}(2005){Heitsch}, {Burkert}, {Hartmann}, {Slyz}, \&
  {Devriendt}}]{heitsch2005}
{Heitsch}, F., {Burkert}, A., {Hartmann}, L.~W., {Slyz}, A.~D., \& {Devriendt},
  J.~E.~G. 2005, \apjl, 633, L113

\bibitem[{{Heitsch} {et~al.}(2006){Heitsch}, {Slyz}, {Devriendt}, {Hartmann},
  \& {Burkert}}]{heitsch2006}
{Heitsch}, F., {Slyz}, A.~D., {Devriendt}, J.~E.~G., {Hartmann}, L.~W., \&
  {Burkert}, A. 2006, \apj, 648, 1052

\bibitem[{{Helmich} {et~al.}(1994){Helmich}, {Jansen}, {de Graauw},
  {Groesbeck}, \& {van Dishoeck}}]{helmich1994}
{Helmich}, F.~P., {Jansen}, D.~J., {de Graauw}, T., {Groesbeck}, T.~D., \& {van
  Dishoeck}, E.~F. 1994, \aap, 283, 626

\bibitem[{{Hildebrand}(1983)}]{hildebrand1983}
{Hildebrand}, R.~H. 1983, \qjras, 24, 267

\bibitem[{{Howell} {et~al.}(1981){Howell}, {McCarthy}, \& {Low}}]{howell1981}
{Howell}, R.~R., {McCarthy}, D.~W., \& {Low}, F.~J. 1981, \apjl, 251, L21

\bibitem[{{Hunter} {et~al.}(2006){Hunter}, {Brogan}, {Megeath}, {Menten},
  {Beuther}, \& {Thorwirth}}]{hunter2006}
{Hunter}, T.~R., {Brogan}, C.~L., {Megeath}, S.~T., {et~al.} 2006, \apj, 649,
  888

\bibitem[{{Imai} {et~al.}(2000){Imai}, {Kameya}, {Sasao}, {Miyoshi}, {Deguchi},
  {Horiuchi}, \& {Asaki}}]{imai2000}
{Imai}, H., {Kameya}, O., {Sasao}, T., {et~al.} 2000, \apj, 538, 751

\bibitem[{{Irvine} {et~al.}(1987){Irvine}, {Goldsmith}, \&
  {Hjalmarson}}]{irvine1987}
{Irvine}, W.~M., {Goldsmith}, P.~F., \& {Hjalmarson}, A. 1987, in Astrophysics
  and Space Science Library, Vol. 134, Interstellar Processes, ed. D.~J.
  {Hollenbach} \& H.~A. {Thronson}, Jr., 561--609

\bibitem[{{Keto}(2002{\natexlab{a}})}]{keto2002a}
{Keto}, E. 2002{\natexlab{a}}, \apj, 568, 754

\bibitem[{{Keto}(2002{\natexlab{b}})}]{keto2002b}
{Keto}, E. 2002{\natexlab{b}}, \apj, 580, 980

\bibitem[{{Keto}(2003)}]{keto2003}
{Keto}, E. 2003, \apj, 599, 1196

\bibitem[{{Klessen} {et~al.}(2005){Klessen}, {Ballesteros-Paredes},
  {V{\'a}zquez-Semadeni}, \& {Dur{\'a}n-Rojas}}]{klessen2005}
{Klessen}, R.~S., {Ballesteros-Paredes}, J., {V{\'a}zquez-Semadeni}, E., \&
  {Dur{\'a}n-Rojas}, C. 2005, \apj, 620, 786

\bibitem[{{Kontinen} {et~al.}(2000){Kontinen}, {Harju}, {Heikkil{\"a}}, \&
  {Haikala}}]{kontinen2000}
{Kontinen}, S., {Harju}, J., {Heikkil{\"a}}, A., \& {Haikala}, L.~K. 2000,
  \aap, 361, 704

\bibitem[{{Krumholz} \& {McKee}(2008)}]{krumholz2008}
{Krumholz}, M.~R. \& {McKee}, C.~F. 2008, \nat, 451, 1082

\bibitem[{{Kumar} {et~al.}(2006){Kumar}, {Keto}, \& {Clerkin}}]{kumar2006}
{Kumar}, M.~S.~N., {Keto}, E., \& {Clerkin}, E. 2006, \aap, 449, 1033

\bibitem[{{Lada} \& {Lada}(2003)}]{lada2003}
{Lada}, C.~J. \& {Lada}, E.~A. 2003, \araa, 41, 57

\bibitem[{{Mac Low} \& {Klessen}(2004)}]{maclow2004}
{Mac Low}, M. \& {Klessen}, R.~S. 2004, Reviews of Modern Physics, 76, 125

\bibitem[{{Megeath} {et~al.}(1996){Megeath}, {Herter}, {Beichman}, {Gautier},
  {Hester}, {Rayner}, \& {Shupe}}]{megeath1996}
{Megeath}, S.~T., {Herter}, T., {Beichman}, C., {et~al.} 1996, \aap, 307, 775

\bibitem[{{Megeath} {et~al.}(2008){Megeath}, {Townsley}, {Oey}, \&
  {Tieftrunk}}]{megeath2008}
{Megeath}, S.~T., {Townsley}, L.~K., {Oey}, M.~S., \& {Tieftrunk}, A.~R. 2008,
  in Handbook of Star Forming Regions, Astronomical Society of the Pacific
  Conference Series,~in press

\bibitem[{{Megeath} {et~al.}(2005){Megeath}, {Wilson}, \&
  {Corbin}}]{megeath2005}
{Megeath}, S.~T., {Wilson}, T.~L., \& {Corbin}, M.~R. 2005, \apjl, 622, L141

\bibitem[{{Mermilliod} \& {Garc{\'{\i}}a}(2001)}]{mermilliod2001}
{Mermilliod}, J.-C. \& {Garc{\'{\i}}a}, B. 2001, in IAU Symposium, 191

\bibitem[{{Mikami} {et~al.}(1992){Mikami}, {Umemoto}, {Yamamoto}, \&
  {Saito}}]{mikami1992}
{Mikami}, H., {Umemoto}, T., {Yamamoto}, S., \& {Saito}, S. 1992, \apjl, 392,
  L87

\bibitem[{{Moore} {et~al.}(2007){Moore}, {Bretherton}, {Fujiyoshi}, {Ridge},
  {Allsopp}, {Hoare}, {Lumsden}, \& {Richer}}]{moore2007}
{Moore}, T.~J.~T., {Bretherton}, D.~E., {Fujiyoshi}, T., {et~al.} 2007, \mnras,
  379, 663

\bibitem[{{Motte} {et~al.}(1998){Motte}, {Andre}, \& {Neri}}]{motte1998}
{Motte}, F., {Andre}, P., \& {Neri}, R. 1998, \aap, 336, 150

\bibitem[{{Neugebauer} {et~al.}(1982){Neugebauer}, {Becklin}, \&
  {Matthews}}]{neugebauer1982}
{Neugebauer}, G., {Becklin}, E.~E., \& {Matthews}, K. 1982, \aj, 87, 395

\bibitem[{{Oey} {et~al.}(2005){Oey}, {Watson}, {Kern}, \& {Walth}}]{oey2005}
{Oey}, M.~S., {Watson}, A.~M., {Kern}, K., \& {Walth}, G.~L. 2005, \aj, 129,
  393

\bibitem[{{Palau} {et~al.}(2007){Palau}, {Estalella}, {Ho}, {Beuther}, \&
  {Beltr{\'a}n}}]{palau2007}
{Palau}, A., {Estalella}, R., {Ho}, P.~T.~P., {Beuther}, H., \& {Beltr{\'a}n},
  M.~T. 2007, \aap, 474, 911

\bibitem[{{Peretto} {et~al.}(2006){Peretto}, {Andr{\'e}}, \&
  {Belloche}}]{peretto2006}
{Peretto}, N., {Andr{\'e}}, P., \& {Belloche}, A. 2006, \aap, 445, 979

\bibitem[{{Peretto} {et~al.}(2007){Peretto}, {Hennebelle}, \&
  {Andr{\'e}}}]{peretto2007}
{Peretto}, N., {Hennebelle}, P., \& {Andr{\'e}}, P. 2007, \aap, 464, 983

\bibitem[{{Preibisch} {et~al.}(1999){Preibisch}, {Balega}, {Hofmann},
  {Weigelt}, \& {Zinnecker}}]{preibisch1999}
{Preibisch}, T., {Balega}, Y., {Hofmann}, K.-H., {Weigelt}, G., \& {Zinnecker},
  H. 1999, New Astronomy, 4, 531

\bibitem[{{Ridge} \& {Moore}(2001)}]{ridge2001}
{Ridge}, N.~A. \& {Moore}, T.~J.~T. 2001, \aap, 378, 495

\bibitem[{{Schilke} {et~al.}(1997){Schilke}, {Walmsley}, {Pineau des Forets},
  \& {Flower}}]{schilke1997a}
{Schilke}, P., {Walmsley}, C.~M., {Pineau des Forets}, G., \& {Flower}, D.~R.
  1997, \aap, 321, 293

\bibitem[{{Sewilo} {et~al.}(2004){Sewilo}, {Churchwell}, {Kurtz}, {Goss}, \&
  {Hofner}}]{sewilo2004}
{Sewilo}, M., {Churchwell}, E., {Kurtz}, S., {Goss}, W.~M., \& {Hofner}, P.
  2004, \apj, 605, 285

\bibitem[{{Tieftrunk} {et~al.}(1997){Tieftrunk}, {Gaume}, {Claussen}, {Wilson},
  \& {Johnston}}]{tieftrunk1997}
{Tieftrunk}, A.~R., {Gaume}, R.~A., {Claussen}, M.~J., {Wilson}, T.~L., \&
  {Johnston}, K.~J. 1997, \aap, 318, 931

\bibitem[{{Tieftrunk} {et~al.}(1998){Tieftrunk}, {Gaume}, \&
  {Wilson}}]{tieftrunk1998}
{Tieftrunk}, A.~R., {Gaume}, R.~A., \& {Wilson}, T.~L. 1998, \aap, 340, 232

\bibitem[{{van der Tak} \& {Menten}(2005)}]{vandertak2005b}
{van der Tak}, F.~F.~S. \& {Menten}, K.~M. 2005, \aap, 437, 947

\bibitem[{{van der Tak} {et~al.}(2005){van der Tak}, {Tuthill}, \&
  {Danchi}}]{vandertak2005a}
{van der Tak}, F.~F.~S., {Tuthill}, P.~G., \& {Danchi}, W.~C. 2005, \aap, 431,
  993

\bibitem[{{van der Tak} {et~al.}(2000){van der Tak}, {van Dishoeck}, {Evans},
  \& {Blake}}]{vandertak2000b}
{van der Tak}, F.~F.~S., {van Dishoeck}, E.~F., {Evans}, II, N.~J., \& {Blake},
  G.~A. 2000, \apj, 537, 283

\bibitem[{{Wilson} {et~al.}(2003){Wilson}, {Boboltz}, {Gaume}, \&
  {Megeath}}]{wilson2003}
{Wilson}, T.~L., {Boboltz}, D.~A., {Gaume}, R.~A., \& {Megeath}, S.~T. 2003,
  \apj, 597, 434

\bibitem[{{Zhang} {et~al.}(1995){Zhang}, {Ho}, {Wright}, \&
  {Wilner}}]{zhang1995}
{Zhang}, Q., {Ho}, P.~T.~P., {Wright}, M.~C.~H., \& {Wilner}, D.~J. 1995,
  \apjl, 451, L71+

\bibitem[{{Ziurys} \& {Friberg}(1987)}]{ziurys1987}
{Ziurys}, L.~M. \& {Friberg}, P. 1987, \apjl, 314, L49

\bibitem[{{Ziurys} {et~al.}(1989){Ziurys}, {Friberg}, \& {Irvine}}]{ziurys1989}
{Ziurys}, L.~M., {Friberg}, P., \& {Irvine}, W.~M. 1989, \apj, 343, 201

\end{thebibliography}

\newpage 
\pagebreak

\newpage
\pagebreak

\end{document}